\documentclass{pasj01}

\Received{$\langle$reception date$\rangle$}
\Accepted{$\langle$acception date$\rangle$}
\Published{$\langle$publication date$\rangle$}

\let\leftroot\relax
\let\uproot\relax

\usepackage{graphicx}
\usepackage{natbib}
\usepackage{ulem} 
\usepackage{physics}
\usepackage[switch,mathlines]{lineno}

\bibpunct[:]{(}{)}{,}{a}{}{,}
\definecolor{ao_en}{rgb}{0.0, 0.5, 0.0}

\makeatletter
\@ifpackageloaded{hyperref}{}{\newcommand{\href}[1]{}}
\makeatother

\begin{document}

\title{Impacts of stellar wind and supernovae on the star cluster formation:
origins of extremely high N/O ratios and multiple stellar populations}
\author{Hajime Fukushima, Hidenobu Yajima}%
\altaffiltext{}{Center for Computational Sciences, University of Tsukuba, Ten-nodai, 1-1-1 Tsukuba, Ibaraki 305-8577, Japan}
\email{fukushima@ccs.tsukuba.ac.jp}


\maketitle

\begin{abstract}

We study metal enrichment originating from stellar wind and supernovae in low-metallicity clouds
by performing three-dimensional radiation hydrodynamics simulations.
We find that metals ejected from stellar wind are accumulated, leading to subsequent star formation in the nitrogen-enriched gas. During this early phase,
the N/O ratios are similar to observed nitrogen-enriched galaxies (${\rm [N/O]}\gtrsim0.5$). 
Then, once supernovae occur, the N/O ratios decrease significantly.
If the duration of star formation is comparable to the timescale of SNe, the mass fraction of nitrogen-enriched stars reaches half the mass of star clusters.
We suggest that the mass of the star cluster needs to exceed $\sim 10^6~M_{\odot}$ to have multiple populations due to stellar wind, considering the condition for massive star cluster formation and the timescales of stellar evolution.

\end{abstract}


\section{INTRODUCTION} \label{introduction}

Globular clusters (GCs) are believed to form in early galaxies, but their formation process is still puzzling.
The candidates of GCs progenitor clusters have been observed as the lensed objects in the era before JWST \citep[e.g.,][]{2017MNRAS.467.4304V, 2019MNRAS.483.3618V}.
Recently, JWST discovered massive and compact star-forming regions in early galaxies. 
They contain the stellar systems with mass and density comparable to that of local GCs \citep[$>10^{5}~M_{\odot}$ and $> 10^3~M_{\odot}{\rm pc^{-3}}$,][]{2010ARA&A..48..431P}.
JWST also reveals a large number of high-z galaxies at $z\gtrsim 10$ \citep[e.g.,][]{2023ApJS..265....5H}.
Among the observed objects, a few galaxies show features of GCs. 
For instance, GN-z11, that is at $z=10.6$ \citep{2023A&A...677A..88B}, has a high N/O abundance ratio \citep{2023MNRAS.523.3516C, 2023arXiv230304179S}.
Recently, \citet{2023ApJ...959..100I} found two other nitrogen-enriched galaxies, and suggested that their abundance patterns might be originated from the CNO cycle. 
Also, \citet{2024A&A...681A..30M} found the nitrogen-enriched galaxy at $z=8.7$.
\citet{2024arXiv240108764T} found a compact star forming region ($\sim 20~{\rm pc}$) at z=6.1 showing a high N/O ratio.
Stars in GCs are typically classified in the first and second populations (1P and 2P). 
Stars in 1P are similar to field stars, but 2P stars have particular light element abundances, e.g., enhancements in He, N, and Na, and depletion in C and O \citep[e.g.,][]{2018ARA&A..56...83B, 2019A&ARv..27....8G}.
The abundance patterns of these galaxies are similar to that of 2P stars  \citep{2023arXiv230304179S}.
These observations imply that the JWST reaches the birthplaces of GCs.

Nitrogen enrichment is enhanced in the CNO cycle \cite[e.g.,][]{2002A&A...381L..25M}.
The stellar wind from Wolf-Rayet stars (WRs) is a candidate process of metal enrichment in GCs \citep{2006PASP..118.1225S}.
In star cluster formation, stellar wind can be mixed with ambient gas, and a fraction of polluted gas gradually increases.
\citet{2021ApJ...922L...3L} performed the simulations of star cluster formation with stellar wind. 
They showed that the mass fraction of gas created from the stellar wind increases as clouds become compact.
In particular, WRs can be nitrogen sources for producing the observed abundance 
 of GN-z11 \citep{2023MNRAS.523.3516C}.
\citet{2024ApJ...962L...6K} showed that metal enrichment of WR stars can reproduce the observed abundance ratio of GN-z11, with two starbursts in a time interval of $\sim 100~M_{\odot}$.
\citet{2024ApJ...962...50W} showed that the N/O ratio becomes comparable to that of the observed nitrogen-enriched galaxies by considering the stellar wind from WRs based on their one-zone calculations.
Nitrogen-enriched gas remains until SNe occurs at the end of the lifetimes of their progenitors because
SNe supplies oxygen-enriched gas, which decreases the N/O ratio significantly.

In addition to the stellar wind, photoionization feedback from massive stars is important to understand the star cluster formation \citep[e.g.,][]{2018ApJ...859...68K, 2020MNRAS.497.3830F}.
In cases of diffuse clouds, the metal-enriched gas from the stellar wind can be evacuated due to the expansion of H{\sc ii} bubbles.
On the other hand, if clouds are massive enough to hold the ionized gas \citep{2021MNRAS.506.5512F}, the metal-enriched gas is used by subsequent star formation, resulting in multiple stellar populations in the same star cluster. 
Thus, we consider the effects of photoinization, stellar wind, SNe
to understand the metal pollution in star cluster formation.

In this paper, we study self-metal enrichment and star formation with different chemical abundances in clouds with masses of $M_{\rm cl} = 10^7$ and $10^8~M_{\odot}$ and surface densities of $\Sigma_{\rm cl} = 400-1500~M_{\odot}{\rm pc}^{-2}$.
Our simulations include radiative, stellar wind, and supernovae (SNe) feedback. 

We organize the rest of the paper as follows.
In Section \ref{numerical_method}, we first describe the numerical method of our simulations.
The main results are presented in Section \ref{result}.
We discuss the implication of our results in Section \ref{Section:discussion}.
Finally, we summarize our results in Section \ref{Section:summary}.

\section{NUMERICAL METHOD} \label{numerical_method}
We make use of \textsc{sfumato-m1} code \citep{2021MNRAS.506.5512F}, the modified version of self-gravitational magnetohydrodynamics code \textsc{sfumato} \citep{2007PASJ...59..905M} that utilize an Eulerian adaptive mesh refinement (AMR) technique.
This code includes the radiation transfer calculation with the M1-closure method \citep{2013MNRAS.436.2188R, 2015ApJ...809..187S}. 
Here, we adopt the approximation of the reduced speed of light with $3 \times 10^{-4}$ times the speed of light. 
Our simulation also includes the chemistry solver \citep{2020ApJ...892L..14S, 2020MNRAS.497..829F} and the sink particle technique to trace the motion of stars \citep{2015ApJ...801...77M}. 
Further details of these methods are shown in \citet{2021MNRAS.506.5512F} and \citet{2022MNRAS.511.3346F}.
We include feedback and metal yield from SNe and stellar wind that induce the metal enrichment of gas in star cluster formation.

\subsection{Model of stellar populations}\label{method_stellar_population}
We assume a sink particle contains several stars stochastically chosen from the modeled initial mass function (IMF) \citep{2022MNRAS.511.3346F}.
In this method, the stellar masses are divided into $150$ bins between $0.1$ and $120~M_{\odot}$ with the Chabrier IMF \citep{2003PASP..115..763C}.
We estimate radiative properties of individual stars in the sink particle considering their ages and masses.
We adopt the \textsc{PARSEC} tracks \citep{2012MNRAS.427..127B, 2014MNRAS.444.2525C,2014MNRAS.445.4287T,2015MNRAS.452.1068C,2017ApJ...835...77M,2019MNRAS.485.5666P,2020MNRAS.498.3283P} for stars with the mass lower than $13~M_{\odot}$.
We use the table of the stellar evolution of effective temperature ($T_{\rm eff}$) and luminosity ($L_*$) calculated by \citet{2018ApJS..237...13L} for more massive stars.
We calculate the emissivity of ionizing photons with the SED models in \citet{2019A&A...621A..85H} considering OB-stars in SMC.

\subsection{Model of stellar wind and SNe}
We adopt the tables of the stellar evolution and metal yield derived by \citet{2018ApJS..237...13L}.
In the recommended model of \citet{2018ApJS..237...13L}, stars more massive than $25~M_{\odot}$ directly collapse into black holes.
Therefore, they contribute to metal enrichment only via stellar wind.
Stars in the mass range of $8-25~M_{\odot}$ induce SNe in their death.
When an SN occurs, we uniformly distribute the ejected metals into cells around the sink particle. 

We divide the stellar evolution into the main sequence and WR star phases.
The stellar model of \citet{2018ApJS..237...13L} gives the mass loss rates and duration time in both phases.
We assume that the wind velocity is 2.6 times the escape velocity for the main-sequence stars \citep{2000A&A...362..295V, 2022ARA&A..60..203V}, and adopt the empirical relation of \citet{2000A&A...360..227N} for the wind velocity of WR stars.
At each timestep, we calculate the ejected metal yield, mass, and momentum rates from stars inside each sink particle. 
In this study, we give feedback to the area with a radius of 4 times the sink radius.
We set the sink radius $r_{\rm sink} = 2 \Delta x$ where $\Delta x$ is the minimum cell length at the maximum refinement level.

SNe occur at the end of lifetimes of stars in the mass range of $8-25~M_{\odot}$. 
Numerical simulations frequently suffer from "overcooling problem" for the SN feedback if the Sedov-Taylor (ST) stage of a supernova remnant is not spatially resolved sufficiently \citep[e.g.,][]{1992ApJ...391..502K}. 
To avoid this problem, we adopt the same strategy used in \citet{2017ApJ...846..133K}.
We calculate the mean gas density ($\bar{n}_{\rm H}$) and total mass inside a spherical shell of the radius $R_{\rm SNR}$ centered around the sink particle where an SN occurs.
Then, we compare the total mass inside the shell $(M_{tot})$ with the expected shell mass at the beginning of the momentum conserving stage $M_{\rm sf} = 1679~M_{\odot} (\bar{n_{\rm H}}/{\rm cm^{-3}})^{-0.26}$ obtained by \citet{2015ApJ...802...99K}.
First, we consider the radius of the supernova remnant (SNR) as $R_{\rm SNR} = 6\Delta x$ where $\Delta x$ is the cell width of the maximum refinement level.
When $M_{\rm tot}/M_{\rm sf} < 0.027$ is satisfied, we can resolve the evolution of ST stages.
Then, we increase $R_{\rm SNR}$ until the inside mass matches $M_{\rm tot}/M_{\rm sf} \geqq 0.027$, and inject the thermal and kinetics energies inside $R_{\rm SNR}$.
We assume that the total energy per SN is $10^{51} ~{\rm erg}$ and the ratios of thermal and kinetic energies are 72 and 28 percent, and inject these energies inside the radius   \citep{2017ApJ...846..133K}.
When we cannot find the injection radius to resolve the ST phase, we inject the momentum into cells around a sink particle.
We adopt the momentum injection rates given by \citet{2015MNRAS.451.2900K}:
\begin{align}
p_{\rm SN} = 3\times 10^5 ~{\rm km~s^{-1}~ M_{\odot}} ~E_{51}^{16/17} n_{\rm H}^{-2/17} Z^{'-0.14}, \label{SNmom}
\end{align}
where $E_{51}$, $n_{\rm H}$ and $Z^{'}$ are SN energy in the unit of $10^{51}~{\rm erg}$, mean gas number density inside the SNR, and the metallicity with a minimum of 0.01 (i.e., $Z^{'} = \max [Z/Z_{\odot}, 0.01])$.

Gas temperature in the stellar wind or SNR exceeds $10^4~{\rm K}$. 
We change the cooling function from our previous model at $T = 3 \times 10^4~{\rm K}$ above which we 
 adopt the one given by \textsc{MAPPINGS V} \citep{2018ascl.soft07005S}.

\subsection{Initial condition} \label{initial_condition}

\begin{table*}
    \caption{Model considered.}
    \label{Tab:model}
    \centering
    \begin{tabular}{|l|c|c|c|c|c|c|}
        \hline
        Model & $M_{\rm cl} (M_{\odot})$ & $R_{\rm cl} ({\rm pc})$ & $\Sigma_{\rm cl} (M_{\odot} {\rm pc^{-2}})$ & $\alpha_{\rm vir}$ & $t_{\rm ff} ({\rm Myr})$ & $M_{\rm cl}/t_{\rm ff} (M_{\odot}/{\rm yr})$ \\
        \hline
        M7R46A2 & $10^7$ & $46$ & $1500$ & $2$ & $1.6$ & $6.1$\\
        M7R64A2 & $10^7$ & $63$ & $800$ & $2$ & $2.6$ & $3.8$ \\
        M7R89A2 & $10^7$ & $89$ & $400$ & $2$ & $4.4$ & $2.3$\\
        M8R200A2 & $10^8$ & $200$ & $800$ & $2$ & $4.7$ & $21$ \\
        M8R200A1 & $10^8$ & $200$ & $800$ & $1$ & $4.7$ &$21$  \\
        M8R200A05 & $10^8$ & $200$ & $800$ & $0.5$ & $4.7$ &$21$  \\
        M8R280A2 & $10^8$ & $280$ & $400$ & $2$ & $7.8$ & $13$\\
      \hline 
    \end{tabular}
    \begin{minipage}{1 \hsize}
    Notes. Column 1: model names, Column 2: cloud masses, Column 3: cloud radii, Column 4: surface densities, Column 5: virial parameters, Column 6: free fall times, Column 7: cloud masses divided by free fall times.
    \end{minipage}
\end{table*}

Table \ref{Tab:model} summarizes the models considered in this paper. 
We set uniform-density spheres with turbulent velocity fields as initial conditions.
We consider clouds with masses $M_{\rm cl} = 10^7, 10^8~M_{\odot}$ and the surface densities $\Sigma_{\rm cl} = 400-1500~M_{\odot}{\rm pc^{-2}}$.
We select these parameters to investigate metal enrichment in star cluster formation, comparing the condition of massive and high-density cluster formation and timescales of stellar evolution.
We further discuss the criterion for the formation of metal-enriched clusters in Section \ref{Subsection:cloud_compactness}. 
As in the previous papers \citep{2020MNRAS.497.3830F, 2021MNRAS.506.5512F}, we adopt the velocity power spectrum of $P(k) \propto k^{-4}$. 
The strength of the turbulent motion is given as the virial parameter defined as
\begin{align}
\alpha_{\rm vir} = \frac{5 \sigma_{0}^2 R_{\rm cl}}{3 G M_{\rm cl}}, \label{eq_alpha_vir}
\end{align}
where $\sigma_0$ is the 3D velocity dispersion.
We adopt the value of $\alpha_{\rm vir} = 2$, with which the turbulent motion balances with the self-gravity of a cloud.
We also perform the simulations with $\alpha_{\rm vir} = 1$ and $0.5$ to investigate the effects of the virial parameters in the cases with $(M_{\rm cl}, R_{\rm cl}) = (10^8~M_{\odot}, 200~{\rm pc})$.
We refine cells to resolve a Jeans length with more than 5 cells.
In each simulation, we set uniform 64 meshes in each direction at the coarsest level, and the maximum refinement level is fixed at $l_{\rm max} = 4$, corresponding to the minimum cell length $\Delta x = 1.56 ~{\rm pc} (R_{\rm cl}/200~{\rm pc})$.
We resolve cells on sink particles with the maximum resolution to simulate the motions of ejected materials.
We set uniform metallicity as $Z=10^{-2} ~Z_{\odot}$ in all cases and assume the solar abundance ratios as the initial state.

\section{RESULTS} \label{result}
We first describe the evolution of metal yield on the star cluster formation in Section \ref{Subsection:metal_yield}. 
In Section \ref{Subsection:cloud_compactness}, we investigate the condition of the formation of nitrogen-enriched stars in massive star clusters. 

\subsection{Metal yield in massive star cluster formation} \label{Subsection:metal_yield}

\begin{figure*}
  \includegraphics[width=160mm]{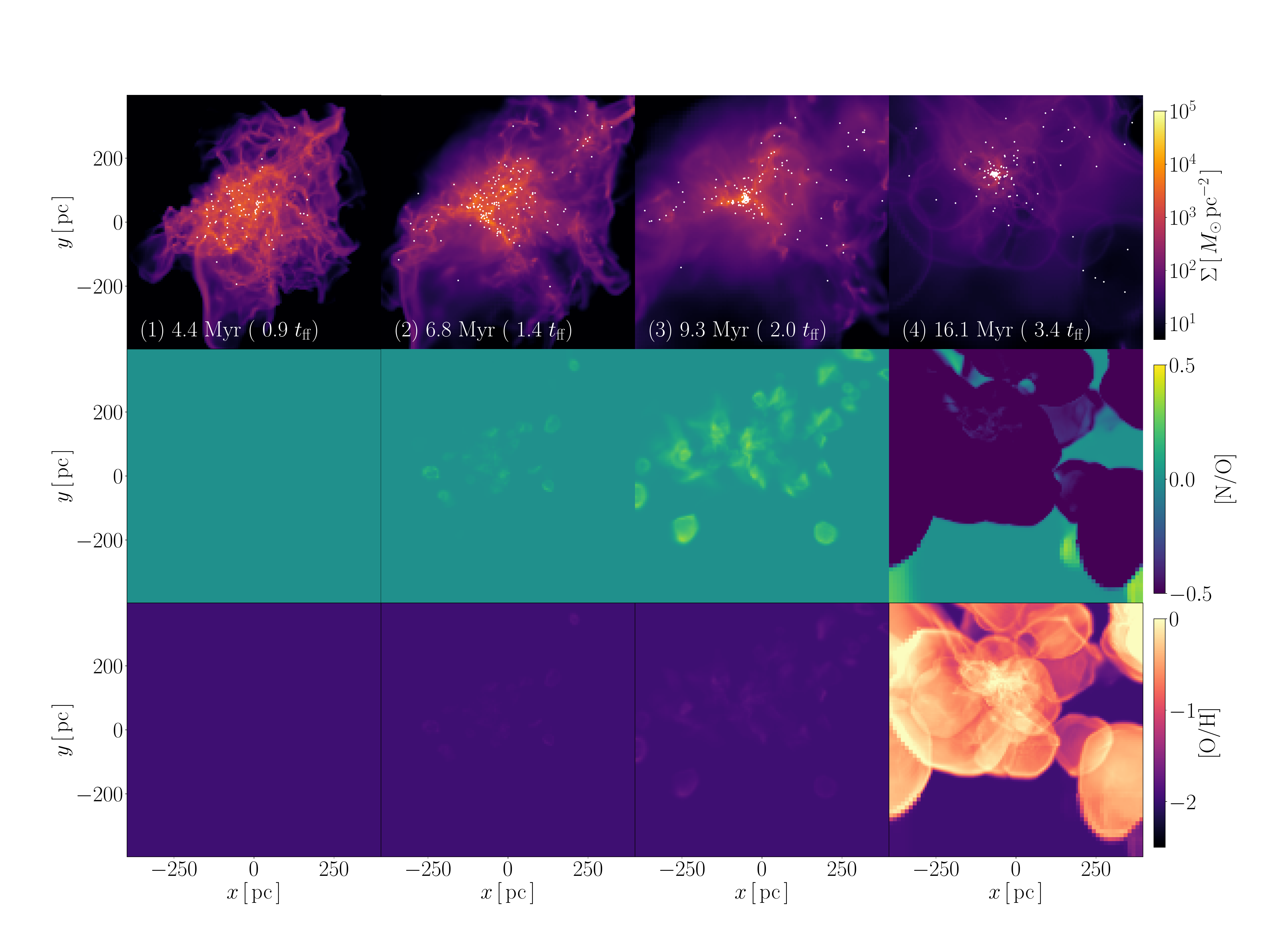}
 \caption{Star cluster formation and metal yield in the case with $M_{\rm cl} = 10^8~M_{\odot}$ and $\Sigma_{\rm cl} = 800~M_{\odot}{\rm pc^{-2}}$. Each panel shows the surface density of the gas ($\Sigma$), the abundance ratio between N and O (${\rm [N/O]}$), and the oxygen abundance to hydrogen ($\rm [O/H]$). 
 The values of ${\rm [N/O]}$ and $\rm [O/H]$ are the mass-weighted averages along the line of sight. 
 The white dots represent the potions of star cluster particles.
 }\label{fig_snap}
\end{figure*}

We first present the case with $(M_{\rm cl}, R_{\rm cl}) = (10^8~M_{\odot}, 200~{\rm pc})$ as the fiducial model. 
Fig. \ref{fig_snap} shows the time evolution of the physical properties of the cloud. 
The star formation mainly occurs after around the free-fall time ($t_{\rm ff}$).
As the star formation proceeds, H{\sc ii} regions expand.
In compact clouds, the gravitational potential of the star cluster overcomes the thermal pressure of H{\sc ii} regions \citep[e.g.,][]{2012ApJ...758L..28B, 2021MNRAS.506.5512F}. 
Therefore, the photoevaporation of ionized gas is suppressed. In that case, a large fraction of the gas is converted into stars, and a massive compact star cluster finally forms.
At $t>10~{\rm Myr}$, massive stars with the mass range of $8-25~\rm M_{\odot}$ induce SNe at the end of their lifetimes that create the structure of spherical shells with hot bubbles.

The second and third rows of Fig.\ref{fig_snap} show the spatial distributions of the N/O and O/H ratios. Here, the notation [N/O] is defined as $\rm log (N/O) - log (N/O)_{\rm \odot}$, where $\rm (N/O)_{\rm \odot}$ is the solar abundance ratio. 
The mass loss rates of the main-sequence stars are not large enough to alter the abundance ratios of the ambient materials.
At $t>5~{\rm Myr}$, stellar ages exceed $3~{\rm Myr}$, and massive stars evolve into WR stars, and hence the nitrogen-enriched regions appear.
The ejected material cannot escape and gradually accumulates around the star cluster.
In the central regions of the star cluster, the nitrogen-enriched gas (${\rm [N/O]} \sim 0.5$) appears at $t=9.3~{\rm Myr}$.
The oxygen abundance is almost constant until SNe occurs.
The first SN occurs when the elapsed time is $10~{\rm Myr}$. 
After that, [O/H] reaches the solar values inside the SNRs as shown in Fig.\ref{fig_snap}-(4).
Thus, [N/O] decreases to less than $-0.5$.

\subsection{Dependence on cloud mass and compactness}\label{Subsection:cloud_compactness}
\begin{figure}
 \begin{center}
  \includegraphics[width=\columnwidth]{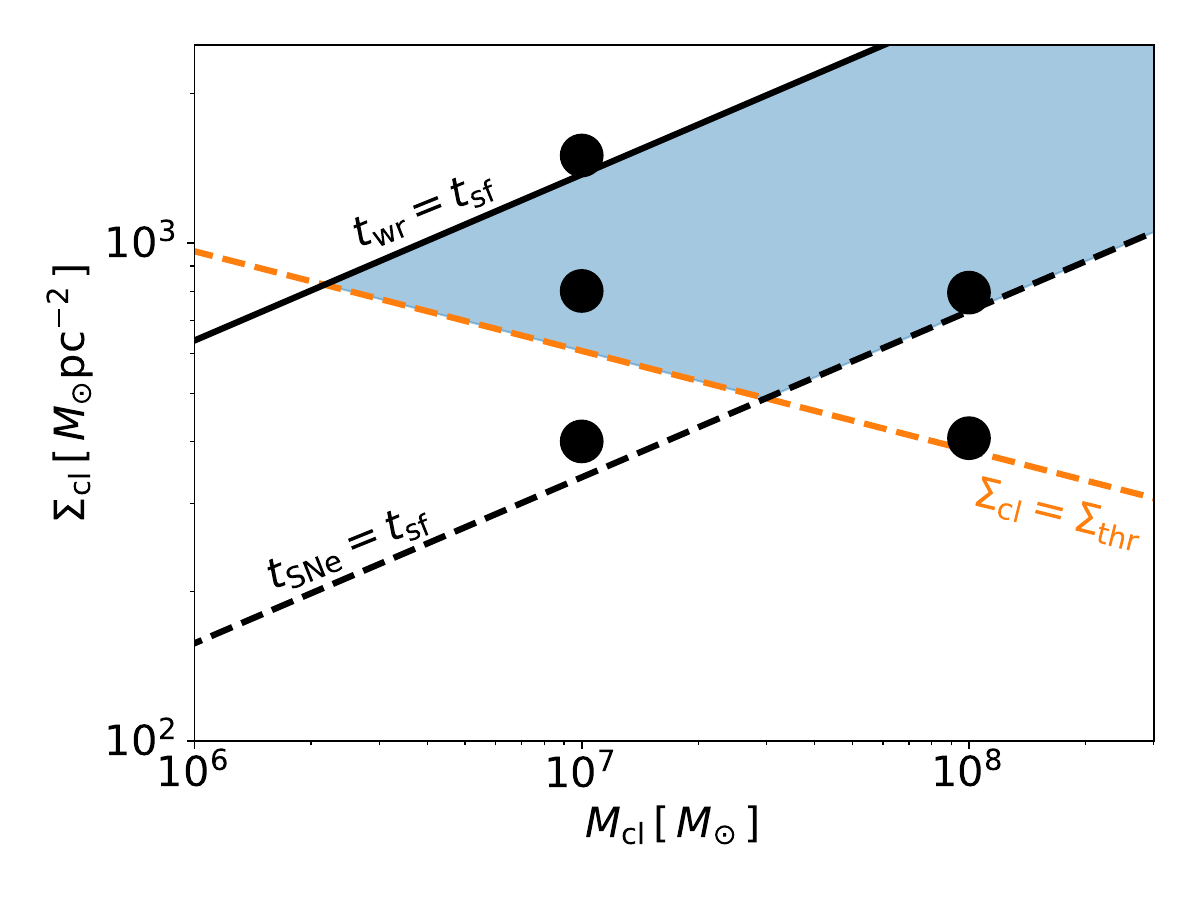}
  \end{center}
      \caption{
      Condition for metal enrichment in star cluster formation. The solid and dashed black lines represent that the duration time of star formation ($t_{\rm sf} \sim 2 ~t_{\rm ff}$) is equal to the timescales of evolution to WR stars ($t_{\rm WR}\sim 3~{\rm Myr}$) and SNe ($t_{\rm SN} \sim 10~{\rm Myr}$).
      The dashed orange line shows the condition for massive star cluster formation (Eq.\ref{eq:sigma_thr}).
    The black dots show the simulation parameters tabulated in Tab.\ref{Tab:model}. 
    The blue shaded region is the parameter space that satisfies the condition for forming the nitrogen-enriched star cluster discussed in Section \ref{section:gas_stellar_metal_yield}.
    }\label{fig_mass_gima}
\end{figure}

In this section, we first consider the conditions of the metal enrichment caused by stellar wind in the massive star cluster formation.
The mass loss rates of low-metallicity main-sequence stars are not high enough to enhance the metal abundance of the ambient gas, while the WR stars are efficient sources of metals. 
The wind from WR stars contains a higher abundance of nitrogen than the solar value.
It induces the nitrogen-enriched state as shown in the middle panels of Fig.\ref{fig_snap}.
At $Z=10^{-2}~Z_{\odot}$, massive stars  with masses larger than $\sim 30~M_{\odot}$ evolve to WR stars $3-5~{\rm Myr}$ after their birth \citep{2018ApJS..237...13L}.
Thus, we set the evolution timescale of WR stars as $t_{\rm WR} \sim 3~{\rm Myr}$.
If the duration time of the star formation is less than $t_{\rm WR}$, the nitrogen enrichment does not affect subsequent star formation.

The SNe also contributes to metal enrichment and is the main oxygen supplier.
We assume that stars in the mass range of $8-25~M_{\odot}$ induce SNe  \citep{2018ApJS..237...13L, 2024ApJ...962...50W}.
The oxygen enrichment from SNe starts after their lifetimes ($t_{\rm SNe}$), which are longer than $10~{\rm Myr}$.
\citet{2010ApJ...713..166B} showed that the inhomogeneous chemical abundance appears in star clusters when the duration time of star formation is longer than the timescale of SNe.

In Fig.\ref{fig_mass_gima}, we summarize the above conditions of metal enrichments.
The duration time of the star formation is typically $\sim 2-3~t_{\rm ff}$ in our simulations. 
We assume that the duration time of the star formation is twice the free-fall time ($t_{\rm ff}$) of the clouds in this figure.
In the region above the solid line, the clouds evolve in a shorter timescale than that of WR stars. 
On the other hand, the metal enrichment of SNe occurs with the cloud parameters below the dashed line.
Therefore, the nitrogen-enriched gas only appears in the parameter region between the solid and dashed lines.
Here, we assume that the timescale of WR stars and SNe are $t_{\rm WR} \sim 3~{\rm Myr}$ and $t_{\rm SNe} \sim 10~{\rm Myr}$. 

\subsubsection{star formation} \label{section:star_formation}

\begin{figure}
  \includegraphics[width=\columnwidth]{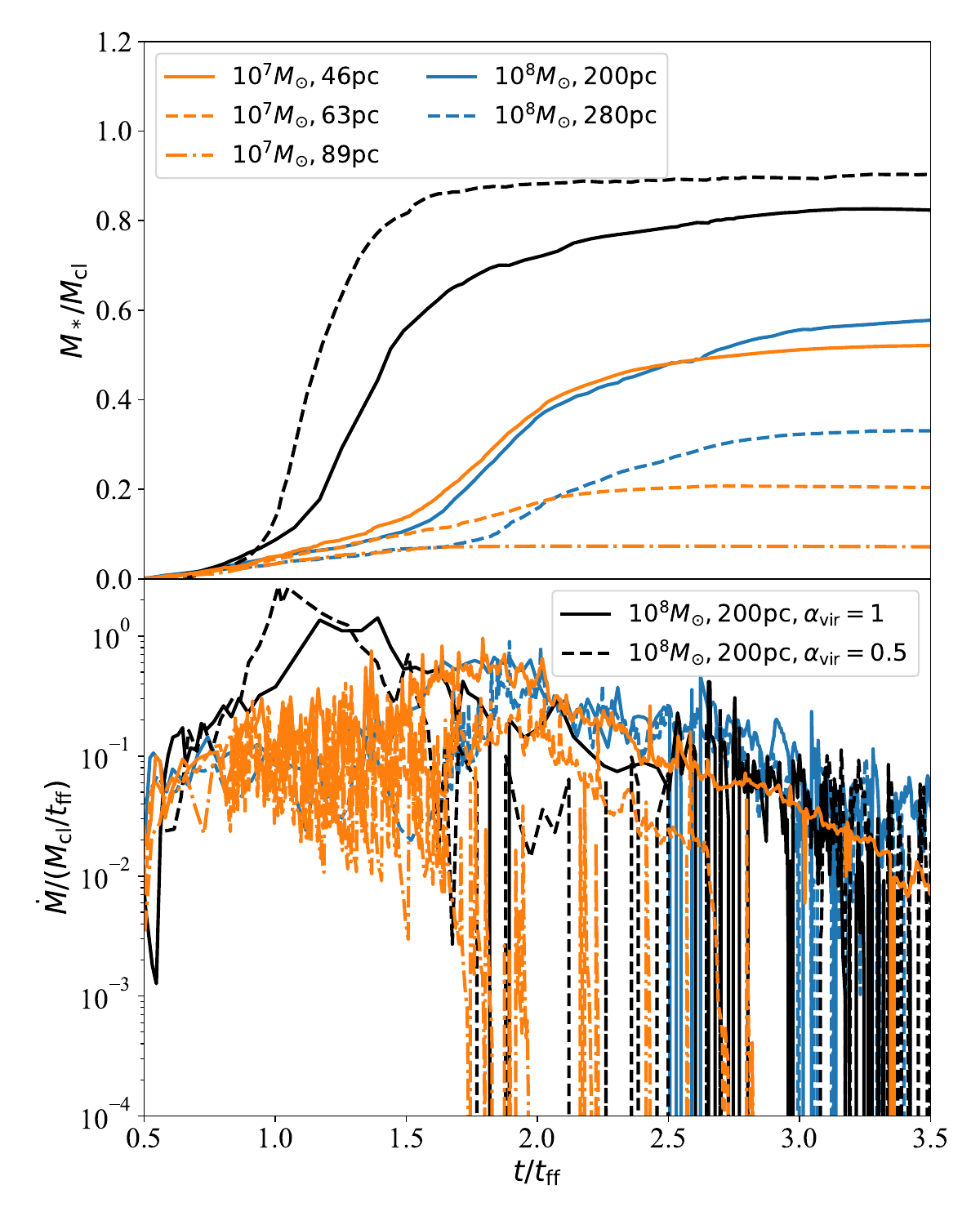}
    \caption{
    Upper panel: The time evolution of the stellar mass in each model. Lower panel: The star formation rates normalized by $(M_{\rm cl}/t_{\rm ff})$.
    Each line shows the models in Tab. \ref{Tab:model}. 
    The blue and orange lines represent the model with $\alpha_{\rm vir} = 2$. 
    The solid (dashed) back line shows the models with $(M_{\rm cl}, R_{\rm cl}) = (10^8~M_{\odot}, 200~{\rm pc})$ and $\alpha_{\rm vir} = 1$ ($\alpha_{\rm vir} = 0.5$). 
    }\label{fig:mass_mdot}
\end{figure}

\begin{figure}
  \includegraphics[width=\columnwidth]{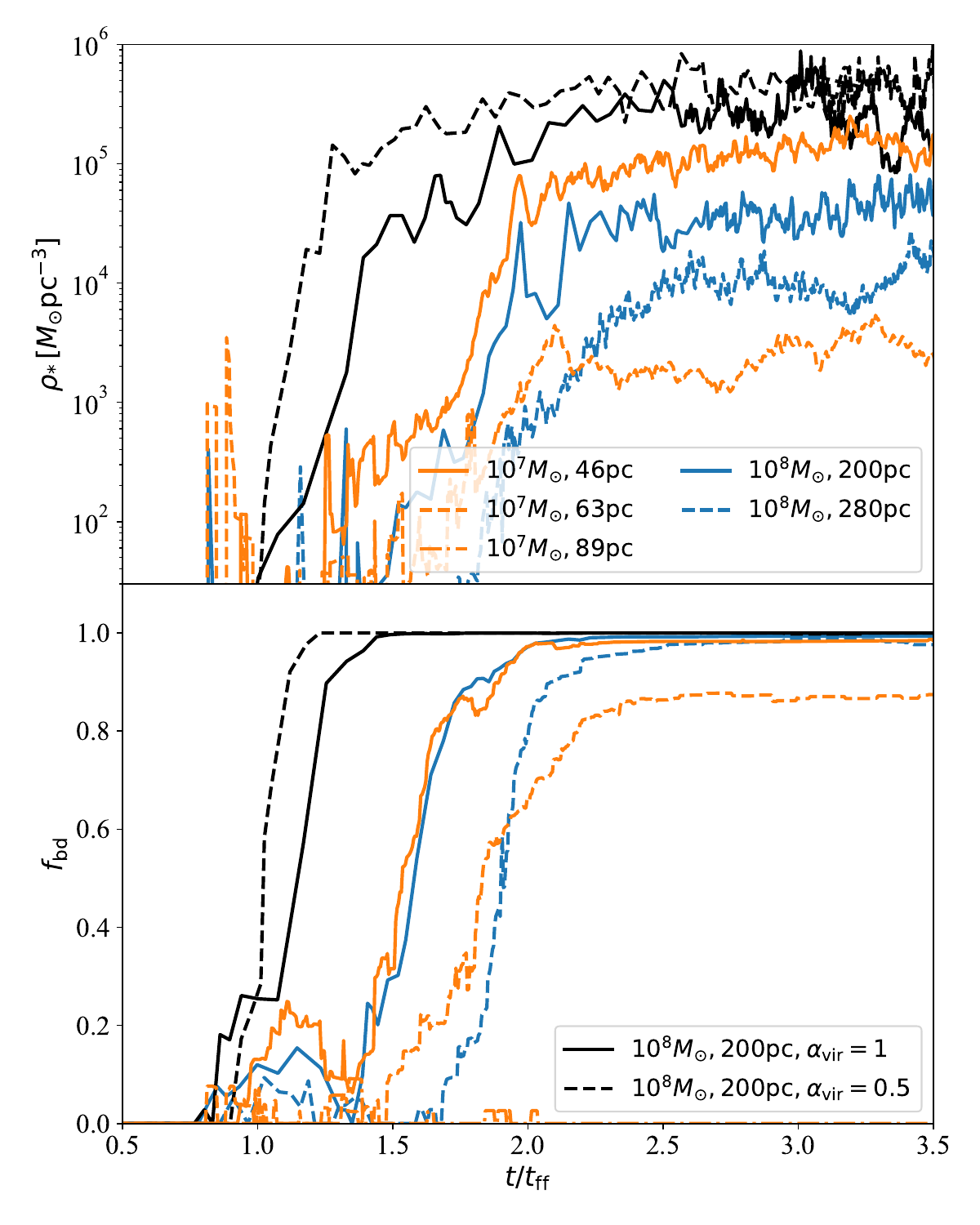}
    \caption{
    Each panel shows the time evolution of stellar densities (upper) and bound fractions (lower). 
    The styles of lines are the same as Fig.\ref{fig:mass_mdot}.
    }\label{fig:bdfrac}
\end{figure}


The star formation efficiencies (SFEs) depend on the cloud surface densities and strength of turbulence motions \citep[e.g.,][]{2018ApJ...859...68K, 2021MNRAS.506.5512F}.
Fig.\ref{fig:mass_mdot} shows the time evolution of total stellar mass and star formation rates.
In models except for $(M_{\rm cl}, R_{\rm cl})=(10^7~M_{\odot}, 89~{\rm pc})$, the star formation efficiencies (SFEs) exceeds 15 percent that is the critical SFE to form bound objects \citep[e.g.,][]{1984ApJ...285..141L, 2001MNRAS.321..699K,2007MNRAS.380.1589B,2017A&A...605A.119S}.
Fig. \ref{fig:bdfrac} shows the time evolution of stellar densities $(\rho_*)$ and bound fractions $f_{\rm bd}$.
The bound fraction is the mass fraction of bound objects to total stellar mass.
We adopt the same strategy to calculate the bound stellar mass as \citet{2017A&A...605A.119S}.
We calculate the stellar densities inside the half-mass radius of bound objects in each simulation.
In these compact clouds, stellar densities and bound fractions rapidly increase in the burst phase of star formation, as shown in Fig. \ref{fig:mass_mdot} and \ref{fig:bdfrac}.
At this epoch, stars start to bind with each other, and their gravitational force accumulates the ambient gas into the star cluster.
Then, the star formation rate (SFR) increases to $0.1-0.5$ times of the cloud masses divided by free fall times ($M_{\rm cl}/t_{\rm ff}$, see also Table \ref{Tab:model}).
For example, the SFR exceeds $10~M_{\odot} {\rm yr^{-1}}$ at this epoch in the case with $(M_{\rm cl}, R_{\rm cl}) = (10^8~M_{\odot},~ 200~{\rm pc})$. 
We note that the star formation rate also depends on the virial parameters. 
In the models with $(M_{\rm cl}, R_{\rm cl}) = (10^8~M_{\odot},~ 200~{\rm pc})$, the maximum SFRs increase to a few $10~M_{\odot} ~{\rm yr^{-1}}$ with the low-virial parameters ($\alpha_{\rm vir} = 0.5$ and $1$).
We show the results with low-virial parameters in Appendix \ref{appendix:low_virial_clouds}.
After that, the stellar density dramatically increase, and the massive and high-density stellar cores appear, as shown in Fig.\ref{fig_snap}.

The cloud surface density for the massive and high-density star cluster formation is given as  \citep{2021MNRAS.506.5512F, 2023MNRAS.524.1422F}  
\begin{align}
    &\Sigma_{\rm cl} > \Sigma_{\rm thr} \nonumber \\
    & = 380~M_{\odot}{\rm pc^{-2}} \left( \frac{\epsilon_{\rm ff}}{0.05} \right)^{2/5} \left( \frac{M_{\rm cl}}{10^8~M_{\odot}} \right)^{-1/5} \left( \frac{T_{i}}{3\times 10^4~{\rm K}} \right)^{28/25}, \label{eq:sigma_thr}     
\end{align}
where $\epsilon_{\rm ff} = \dot M/(M_{\rm cl}/t_{\rm ff})$, and $T_{\rm i}$ is temperature in H{\sc ii} regions.
Here, we adopt the typical temperature of H{\sc ii} regions $T_{\rm i} = 3 \times 10^4~{\rm K}$ at $Z=10^{-2}~Z_{\odot}$ \citep{2020MNRAS.497.3830F, 2023arXiv231213339C}.
In Fig.\ref{fig_mass_gima}, the dashed orange represents the condition of Eq.\eqref{eq:sigma_thr}.
When clouds satisfy these conditions, the ejected material from stellar wind accumulates in the star cluster and is recycled for the subsequent star formation.
In the diffuse cloud model with $(M_{\rm cl}, R_{\rm cl})=(10^7~M_{\odot}, 89~{\rm pc})$, the photoionization feedback efficiently disrupts the cloud, leading to escape of the ejected material.

In the models with $\alpha_{\rm vir} = 2$, the initial turbulent motions decay, and the star formation mainly occurs after the elapsed times $t \sim 1-2~t_{\rm ff}$ via the collapse of the gas clump gravitationally.
In cases with lower virial parameters, star formation proceeds more quickly because the turbulent portion cannot hinder the collapse of the gas.
Here, we additionally simulate the models with $(M_{\rm cl}, R_{\rm cl}) = (10^8~M_{\odot}, 200~{\rm pc})$, adopting the virial paramaters as $\alpha_{\rm vir} = 1$ and $0.5$. 
The star formation rate rapidly increases at $t \sim 1-1.5 ~t_{\rm ff}$, and then the star formation is almost quenched at $t\sim 2 ~t_{\rm ff}$ in both cases. 
Also, rapid star formation induces effective conversion from gas to star.
In the case with $\alpha_{\rm vir} = 0.5$, the SFE exceeds 0.8 finally.

\subsubsection{Gas and stellar metal yield} \label{section:gas_stellar_metal_yield}

\begin{figure*}
    
\begin{minipage}[b]{0.49\linewidth}
\centering
  \includegraphics[width=\columnwidth]{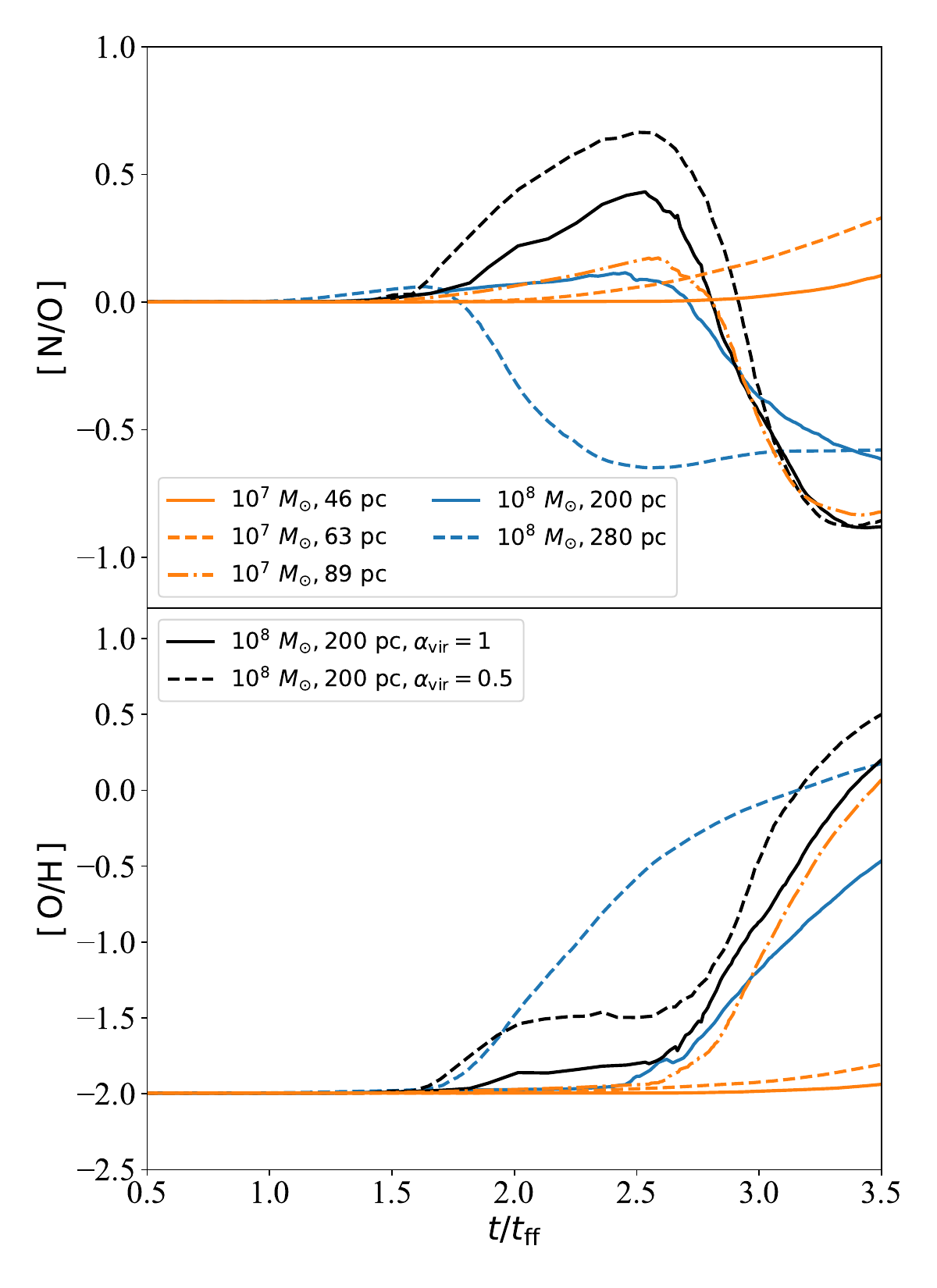}
\end{minipage}
\begin{minipage}[b]{0.49\linewidth}
\centering
  \includegraphics[width=\columnwidth]{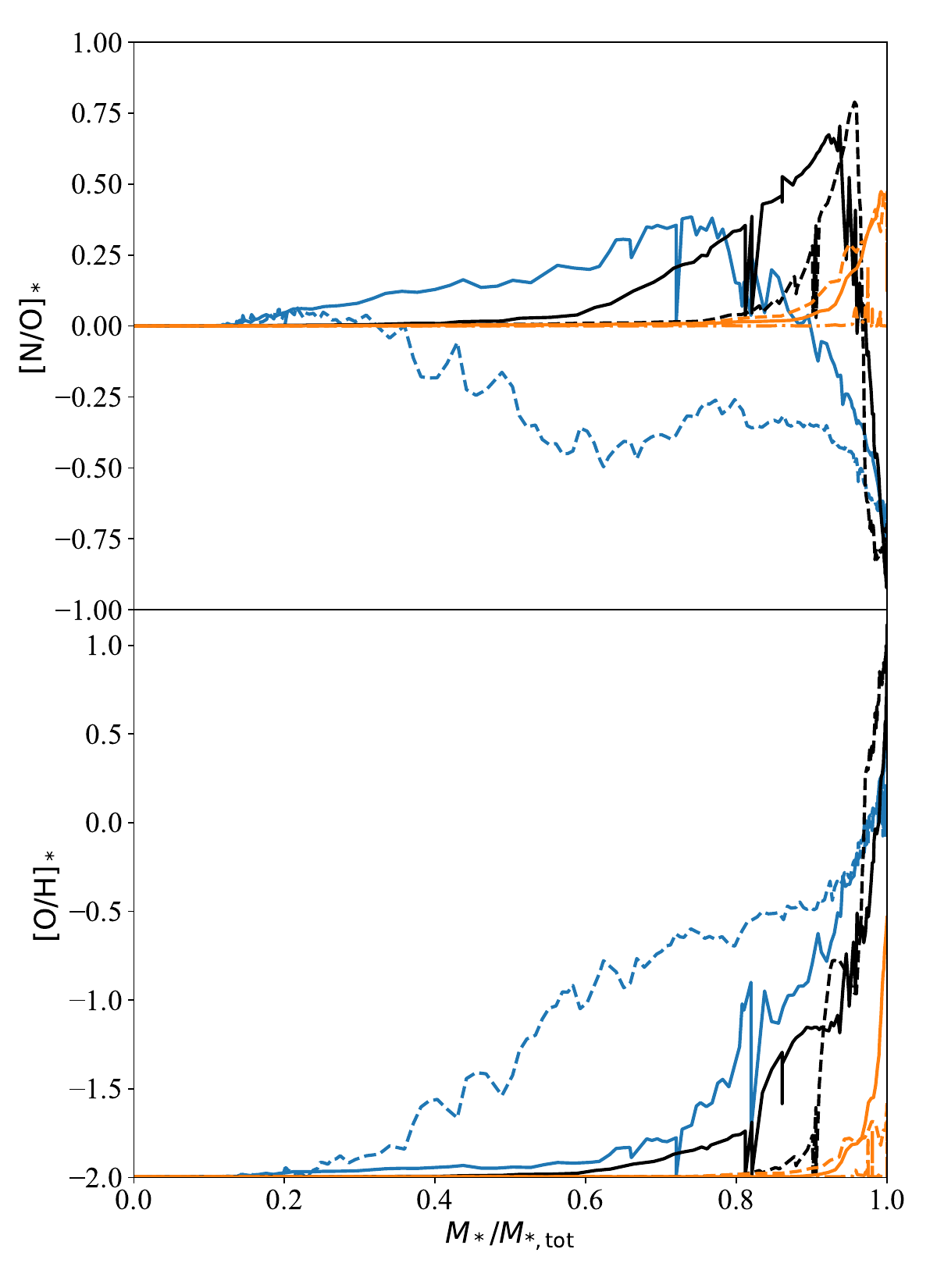}
\end{minipage}
\caption{ Left panel: The time evolution of the mass-weighted averages of $\rm [N/O]$ and $\rm [O/H]$ of gas in computational domain. 
Right panel: The histories of the stellar mass-weighted averages of [N/O] and [O/H]  as the function of $M_*/M_{*, \rm tot}$ where $M_*$ is the masses of the star clusters at the birth time of each star, and $M_{*, \rm tot}$ is the total stellar mass at the end of the simulations. 
   The styles of lines are same as Fig.\ref{fig_mass_gima}.}\label{fig:stellar_abundance}
\end{figure*}
Metal abundances change significantly with time due to the stellar wind and SNe.
The left panel of Fig.\ref{fig:stellar_abundance} shows the time evolution of N/O and O/H ratios of the gas.
Stellar metal abundances should be different from the initial state of the gas clouds if they form in the metal-polluted gas.
The right panel of Fig.\ref{fig:stellar_abundance} represents the metal abundances of the newborn stars. 
As discussed in Section \ref{section:star_formation}, the ejected materials are rapidly evaporated from the star cluster if clouds do not satisfy the condition of Equation \eqref{eq:sigma_thr}. 
Therefore, the metal abundances are almost constant in the case with $(M_{\rm cl}, R_{\rm cl}) = (10^8~M_{\odot}, 280~{\rm pc})$.
On the other hand, in massive star cluster formation, the ejected materials are accumulated around the star cluster and recycled into the star formation.
The evolution of the metal abundances can be classified as in the following (i), (ii), and (iii), depending on whether the duration time of the star formation ($t_{\rm sf}\sim 2~t_{\rm ff}$) is longer than the timescales of WR stars and SNe ($t_{\rm wr}$ and $t_{\rm SNe}$) or not. 

\begin{itemize}
\item[(i)]$t_{\rm sf} < t_{\rm wr},t_{\rm SNe}$: In this case, the star formation rapidly proceeds. Even stellar wind from WR stars cannot contribute to metal enrichment. The model with $(M_{\rm cl}, R_{\rm cl}) = (10^7~M_{\odot}, 46~{\rm pc})$ corresponds to this case. The stellar metal abundance slightly changes in the final phase, but the metal abundances of most stars and gas are almost constant.

\item[(ii)] $t_{\rm sf} > t_{\rm wr}, t_{\rm sf} < t_{\rm SNe}$: Massive stars evolve to WR stars during star formation. The nitrogen-enriched gas appears around the star cluster as shown in Fig.\ref{fig_snap}. 
Then, the nitrogen-enriched (2P) stars are born from the metal-enriched gas.
The fraction of 2P stars also depends on the timescale of the star formation.
The mass fraction of 2P stars increases as the duration time of the star formation becomes longer.
For example, more than half of stars are born as 2P stars in the case with $(M_{\rm cl}, R_{\rm cl}, \alpha_{\rm vir}) = (10^8~M_{\odot}, 200~{\rm pc}, 2)$ in which the duration time is comparable to the timescale of SNe.
In the cases with $\alpha_{\rm vir} = 1$ and $0.5$, the maximum values of [N/O] exceeds 0.5.
These values are comparable to the observed high-redshift galaxies \citep{2023ApJ...959..100I}. 

\item[(iii)]$t_{\rm sf} > t_{\rm wr}, t_{\rm sf} > t_{\rm SNe}$: 
SNe provide oxygen-enriched gas during the star formation. Therefore, the model with $(M_{\rm cl}, R_{\rm cl}) = (10^8~M_{\odot}, 280~{\rm pc})$ shows that most stars form in the oxygen-enriched stars, resulting in the lower [N/O] ratio.
\end{itemize}

As summarized above, the nitrogen-enriched star clusters appear when the condition of (ii) is satisfied. 
The blue-shaded region in Fig. \ref{fig_mass_gima} corresponds to the host clouds of the nitrogen-enriched star clusters.
We suggest that clouds more massive than $\sim 10^7~M_{\odot}$ can form star clusters with multiple populations (the resultant stellar mass is a few $\times 10^{6}~\rm M_{\odot}$), including non-negligible 2P stars.

\section{DISCUSSION} \label{Section:discussion}

In this paper, we study the effects of the stellar wind and SNe on the N/O and O/H ratios in star cluster formation.
Recent observations by JWST have discovered nitrogen-enriched galaxies \citep[e.g.,][]{2023ApJ...959..100I}.
GN-z11 shows the strong nitrogen emission \citep{2023A&A...677A..88B} and the very high nitrogen-oxygen abundance ratio ${\rm [N/O]}\sim 0.5$ \citep{2023MNRAS.523.3516C, 2023arXiv230304179S}.
We suggest that such nitrogen enrichment occurs when the duration time of star formation is longer than the timescales of WR stars. 
Recently, \citet{2024arXiv240108764T} discovered nitrogen-enriched star-forming regions in a few $10$ pc scale in high-z galaxies at $z=6.1$. 
In this region, the stellar mass is $\sim 10^8~M_{\odot}$, and the stellar density is $> 10^4~M_{\odot}{\rm pc^{-2}}$.
If the host gas cloud had a similar density and mass, it is assigned to the category (ii) of our models in which [N/O] becomes high due to the stellar wind from WR stars.

Massive and compact star clusters in high-redshift galaxies are plausible candidates for current globular clusters \citep[e.g.,][]{2014CQGra..31x4006K, 2016ApJ...831..204R, 2018MNRAS.475.4252A}.
Most globular clusters show multiple populations (MPs) of stars \citep[e.g.][, and the reference therein]{2018ARA&A..56...83B, 2019A&ARv..27....8G}.
Considering the conditions presented in the previous section, massive star clusters with the mass $\gtrsim 2 \times 10^{6}~\rm M_{\odot}$ ($M_{\rm cl} \times \rm SFE$) can consist of stars with MPs.
These results are consistent with the fact that the young massive star clusters have no MPs  \citep[$\lesssim 10^6~M_{\odot}$ and $\lesssim 2~{\rm Gyr}$,][]{2018ARA&A..56...83B}.
Enriched stars are born after WR stars appear, and their fractions depend on the masses of host clouds.
Observations of GCs suggest that the fraction of enriched stars increases in more massive star clusters \citep{2017MNRAS.464.3636M}. 
Our results with WRs nicely match the observed trend.

Recently, JWST observations reveal that high-z galaxies contain massive and dense star-forming clumps \citep[e.g.,][]{2023ApJ...945...53V, 2024arXiv240103224A, 2024arXiv240218543F}.
Discovered clumps are massive and compact enough to form massive star clusters ($M_{\rm cl}>10^6~M_{\odot}$ and $\Sigma_{\rm cl} > 10^3~M_{\odot}{\rm pc^{-2}}$), and they dominate the luminosity of the galaxies \citep{2024arXiv240218543F}.
Efficient star formation occurs in massive star cluster formation because photoionization cannot suppress star formation \citep{2021MNRAS.506.5512F}.
Enhancement of star formation efficiencies in massive star cluster formation is a candidate mechanism for the high abundance of bright high-z galaxies \citep{2022ApJ...938L..10I, 2023ApJS..265....5H, 2023MNRAS.525.4832Y}. 
\citet{2023MNRAS.523.3201D} also discussed the feedback-free starburst causing the high abundance of bright galaxies. 
We will apply our results to studies on early galaxy formation and reveal the role of massive star cluster formation in the high abundance of bright galaxies at $z>10$.

Here, we focus only on the stellar wind from WRs and SNe as sources of metal enrichment. On the other hand, other scenarios can be considered to induce the high [N/O] ratios.
\citet{2023A&A...673L...7C} argue that the wind from supermassive stars (SMS) can reproduce the observed abundance ratio of GN-z11 and in GCs considering the scenario in \citet{2018MNRAS.478.2461G}. 
A SMS can form via stellar collision processes in dense star clusters \citep{2002ApJ...576..899P, 2014MNRAS.439.1003F, 2016MNRAS.457.2423Y}.
The mass of SMS sensitively depends on the initial properties of star clusters \citep[e.g.,][]{2021PASJ...73.1074F, 2022MNRAS.514...43F, 2023arXiv231206509P}.

The metal yield changes with the shape of the IMF.
In this work, we have assumed the Chabrier IMF with the mass range of $0.1-120~M_{\odot}$. 
The previous studies showed that the stellar IMF at low-metallicity is top-heavy due to lack of metal cooling  \citep[e.g.,][]{2005ApJ...626..627O,2021MNRAS.508.4175C,2023arXiv231213339C}, and found the implication of top-heavy IMF in globular cluster formation \citep[e.g.,][]{2012MNRAS.422.2246M, 2022MNRAS.516.3342W}.
In addition, stars more massive than $200~M_{\odot}$ can form even in low-metallicity environments \citep{2020MNRAS.497..829F}.
We will consider the mass growth of individual stars in our future simulations of the star cluster formation.

In this work, we assume that massive stars with masses larger than $25~M_{\odot}$ do not induce SNe, and they directly collapse into black holes.
If SNe occur even for such massive stars, the timescale of SNe becomes shorter, and
 nitrogen-enriched star clusters form only in more compact gas clouds, as shown in Fig. \ref{fig_mass_gima}.
The threshold mass for SNe depends on metallicity, rotational velocity, and stellar evolution model \citep[e.g.,][]{2003ApJ...591..288H}. 
The previous studies on stellar evolution show that 
the threshold masses are in
the mass range of $20~M_{\odot}-30~M_{\odot}$ \citep{2016ApJ...821...38S}.
Therefore, our results on the formation conditions of nitrogen-enriched star clusters do not change significantly.

\section{SUMMARY}\label{Section:summary}

We have performed 3D RHD simulations of the massive star cluster formation with $M_* > 10^6~M_{\odot}$, including metal yield from stellar wind and SNe.
Our simulations cover the various cloud masses and surface densities, $M_{\rm cl}=10^7-10^8~M_{\odot}$ and $\Sigma_{\rm cl} = 400-1500~M_{\odot}{\rm pc^{-2}}$. 
Our findings are summarized as follows:

\begin{itemize}
    \item [(i)] Material ejected via stellar wind and SNe is recycled to star formation only in the massive and compact star cluster formation. 
    Before SNe occurs, WR stars enhance the N/O abundance ratio.
    Combining the condition for compact and massive star cluster formation as $ \Sigma_{\rm cl} \gtrsim 380~M_{\odot}{\rm pc}^{-2} (M_{\rm cl}/10^8~M_{\odot})^{-1/5}$,  nitrogen-enriched star clusters forms only in star clusters more massive than $10^6~M_{\odot}$.
    \item[(ii)] The SNe can also contribute to the metal enrichment of gas and stars when the duration time scale of star formation is much longer than the timescale of SNe ($\sim 10~{\rm Myr}$). In such a case, most stars are oxygen-enriched, resulting in lower N/O ratios.
    \item [(iii)] 
    Star formation proceeds rapidly when a cloud virial parameter is low.
    For the virial parameters of $0.5$ and $1.0$, the star formation rate is higher than a few $10~M_{\odot}{\rm yr}^{-1}$ and the N/O abundance ratio exceeds 0.5.
\end{itemize}

Massive and compact clumps discovered in high-z galaxies are candidates for formation sites of globular clusters.
We show that star formation proceeds efficiently in such clumps, and these results are consistent with that discovered clumps dominate the luminosity of host galaxies.
We suggested that star formation in massive star clusters has the potential to cause a high abundance of bright galaxies at $z>10$, and multiple populations in GCs are relics of starbursts in early galaxies.

\section{ACKNOWLEDGEMENTS}
The authors thank Masami Ouchi, Yuki Isobe and Kuria Watanabe for fruitful discussion. 
The numerical simulations were performed on the Cary XC50 (Aterui II) at the Center for Computational Astrophysics of National Astronomical Observertory of Japan and Yukawa-21 at Yukawa Institute for Theoretical Physics in Kyoto University.
This work is supported in part by the MEXT/JSPS KAKENHI graph numbers 23K13139 (HF), 21H04489, 21H01133, 22H00149 (HY), JST FOREST Program, Grant Number JP-MJFR202Z (HY).

\begin{figure*}
 \begin{center}
  \includegraphics[width=160mm]{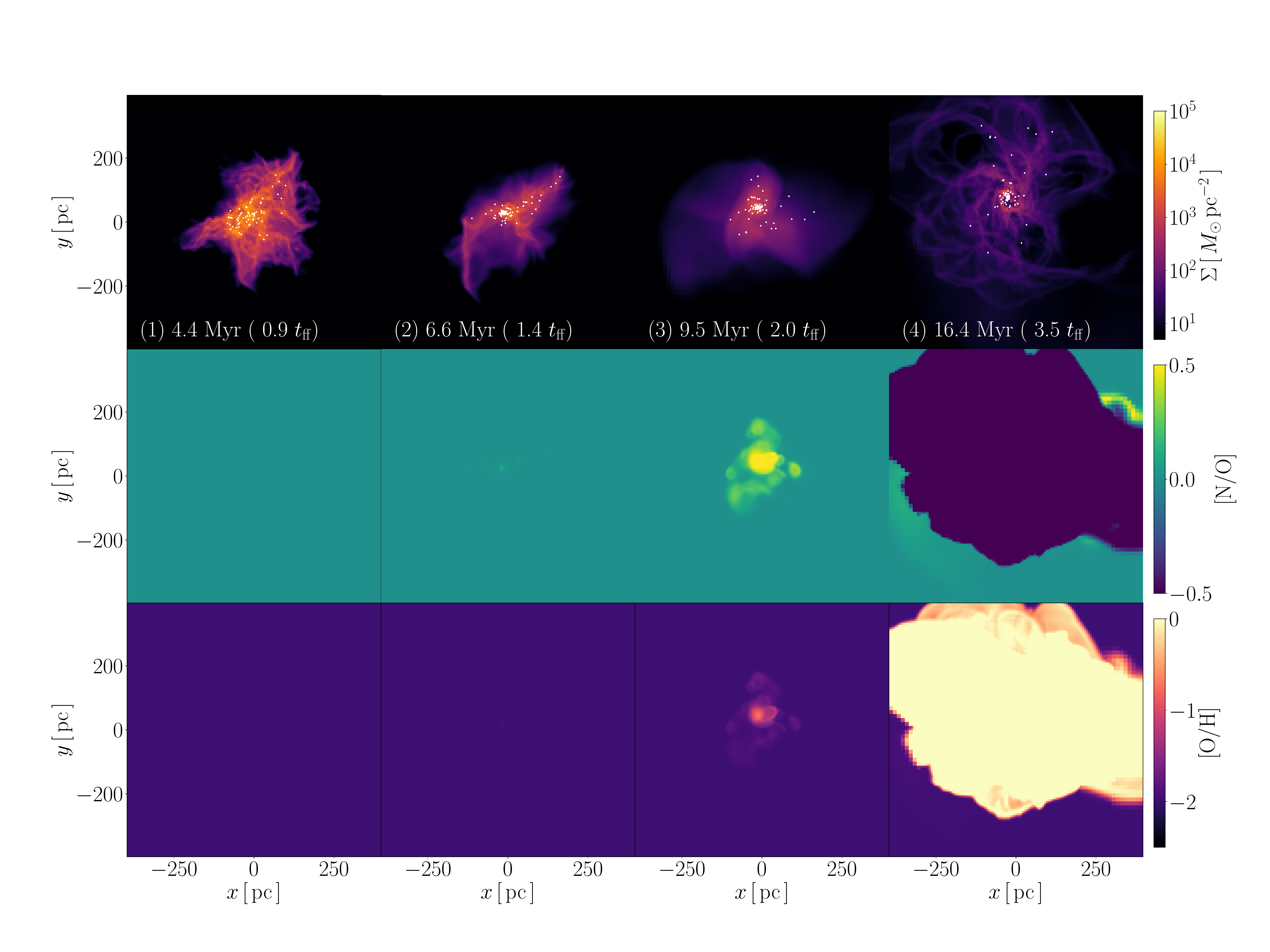}
 \caption{Same as Fig. \ref{fig_snap}, but for the case with $\alpha_{\rm vir} = 0.5$.}\label{fig_snap2}
 \end{center}
\end{figure*}

\bibliographystyle{apj}

\appendix
\section{Low-virial clouds} \label{appendix:low_virial_clouds}
As discussed in Section \ref{section:gas_stellar_metal_yield}, star formation rates and metal abundances depend on the virial parameters of clouds. 
Fig \ref{fig_snap2} shows the model with the same as Fig \ref{fig_snap} but for $\alpha_{\rm vir} = 0.5$.
With low-virial parameters, gas falls on the central region of a cloud. 
Then, the bursty star formation occurs as shown in Fig. \ref{fig:mass_mdot}. 
At this epoch, the star formation rates increase to a few $10~M_{\odot} ~{\rm yr^{-1}}$ in both cases.
The star cluster is also more massive and compact than that with higher virial parameters.
The amount of ejected materials due to stellar wind increases, and gravitational force attracts more ambient gas.
The compact nitrogen-enriched regions appear at $t \sim 9.5~{\rm Myr}$ and its N/O ratio exceeds 0.5. 
These values are similar to observed high redshift galaxies \citep{2023arXiv230304179S, 2023ApJ...959..100I, 2024arXiv240108764T}.
After SNe begins, the gas gradually evaporates from the star cluster, and the gas is oxygen-enriched in SNRs.
\end{document}